\documentclass[12pt,a4j]{article} 
\makeatletter
\newdimen{\eqarcolsep}
\usepackage[dvips]{color,graphicx} 
\usepackage{amsmath} 
\usepackage{amsthm} 
\usepackage{bm} 
\usepackage{fancybox} 
\usepackage{ascmac} 
 \usepackage{amssymb}

\makeatletter 
\begin{document}

{\large \bf Non-self-adjoint hamiltonians defined by generalized Riesz bases} 

\ \\
\begin{center}
Hiroshi Inoue and Mayumi Takakura\\
\end{center}

\ \\
\begin{abstract} 
In \cite{b-i-t}, F. Bagarello, A. Inoue and C. Trapani investigated some operators defined by Riesz bases. These operators connect with ${\it quasi}$-${\it hermitian \; quantum \; mechanics}$, and its relatives. In this paper, we change the frameworks of these operators, and then almost results obtained in \cite{b-i-t} become trivial. Furthermore, we introduce a notion of generalized Riesz bases which is a generalization of Riesz bases and investigate some operators defined by generalized Riesz bases. \\
\end{abstract}

\section{Introduction}
In \cite{b-i-t}, F. Bagarello, A. Inoue and C. Trapani investigated the operators
\begin{eqnarray}
H_{\phi , \psi}^{\bm{\alpha}} 
&=& \sum_{n=0}^{\infty} \alpha_{n} \phi_{n} \otimes \bar{\psi}_{n}, \nonumber \\
A_{\phi , \psi}^{\bm{\alpha}} 
&=& \sum_{n=0}^{\infty} \alpha_{n+1} \phi_{n} \otimes \bar{\psi}_{n+1}, \nonumber \\
B_{\phi , \psi}^{\bm{\alpha}}
&=& \sum_{n=0}^{\infty} \alpha_{n+1} \phi_{n+1} \otimes \bar{\psi}_{n}, \nonumber 
\end{eqnarray}
where $\{ \phi_{n} \}$ is a Riesz basis, that is, there exist a bounded operator $T$ on a Hilbert space ${\cal H}$ with bounded inverse and an ONB $\{ e_{n} \}$ in ${\cal H}$ such that $ \phi_{n} =T e_{n}, \; n= 0,1,2, \cdots$, and $\psi_{n} \equiv \left( T^{-1} \right)^{\ast} e_{n}, \; n= 0,1,2, \cdots$. Then $\{ \phi_{n} \}$ is biorthogonal to $\{ \psi_{n} \}$, that is, $ \left( \phi_{n} | \psi_{m} \right) = \delta_{nm}$ $( n,m = 0,1,2, \cdots )$. These operators connect with ${\it quasi}$-${\it hermitian \; quantum \; mechanics}$, and its relatives \cite{mostafazadeh, bagarello11, bagarello2013}.

In this paper, we generalize some of the results for the above operators defined by Riesz bases. In details, we consider the case that an operator $T$ and its inverse $T^{-1}$ in definition of Riesz bases are not necessary bounded. And we define and study the following operators
\begin{eqnarray}
H_{\bm{\alpha}} 
&\equiv& T \left( \sum_{n=0}^{\infty} \alpha_{n} e_{n} \otimes \bar{e}_{n} \right) T^{-1} , \nonumber \\
A_{\bm{\alpha}}
&\equiv& T \left( \sum_{n=0}^{\infty} \alpha_{n+1} e_{n} \otimes \bar{e}_{n+1} \right) T^{-1} , \nonumber \\
B_{\bm{\alpha}}
&\equiv& T \left( \sum_{n=0}^{\infty} \alpha_{n+1} e_{n+1} \otimes \bar{e}_{n} \right) T^{-1} , \nonumber
\end{eqnarray}
instead of the operators $H_{\phi , \psi}^{\bm{\alpha}} , \; A_{\phi , \psi}^{\bm{\alpha}}$ and $B_{\phi , \psi}^{\bm{\alpha}}$. In fact, in case that $T$ and $T^{-1}$ are bounded, these operators $H_{\bm{\alpha}}, \; A_{\bm{\alpha}}$ and $B_{\bm{\alpha}}$ coincide with the operators $H_{\phi , \psi}^{\bm{\alpha}} , \; A_{\phi , \psi}^{\bm{\alpha}}$ and $B_{\phi , \psi}^{\bm{\alpha}}$, respectively and almost results in \cite{b-i-t} are easily proved using $H_{\bm{\alpha}}$, $A_{\bm{\alpha}}$ and $B_{\bm{\alpha}}$ instead of $H_{\phi , \psi}^{\bm{\alpha}} , \; A_{\phi , \psi}^{\bm{\alpha}}$ and $B_{\phi , \psi}^{\bm{\alpha}}$. Furthermore, though these operators do not equal in general, it seems to be much simpler and better for calculations and clearness to consider the operators  $H_{\bm{\alpha}}, \; A_{\bm{\alpha}}$ and $B_{\bm{\alpha}}$ than the operators $H_{\phi , \psi}^{\bm{\alpha}} , \; A_{\phi , \psi}^{\bm{\alpha}}$ and $B_{\phi , \psi}^{\bm{\alpha}}$. So, we investigate the operators $H_{\bm{\alpha}}, \; A_{\bm{\alpha}}$ and $B_{\bm{\alpha}}$ in details.

This article is organized as follows. In Section 2, we introduce a notion of generalized Riesz bases which is a generalization of Riesz bases and investigate some operators defined by generalized Riesz bases. In particular we characterize domains and adjoints and compare $H_{\bm{\alpha}}$ with $H_{\phi , \psi}^{\bm{\alpha}}$. In Section 3, we construct and study generalized rising and lowering operators $A_{\bm{\alpha}}$ and $B_{\bm{\alpha}}$ defined by generalized Riesz bases. 

\section{Some operators defined by generalized Riesz bases}
Let $T$ be a densely defined closed linear operator in ${\cal H}$ with a densely defined inverse. Then $T^{\ast}$ has a densely defined inverse and $(T^{\ast})^{-1}=(T^{-1})^{\ast}$. Indeed, since the range $R(T)$ of $T$ is dense in ${\cal H}$, it follows that $ker  T^{\ast}= \{ 0 \}$ and $R(T^{\ast})$ is dense in ${\cal H}$. Furthermore, $( \xi | \eta)=(T^{-1}T \xi | \eta)=(T \xi | (T^{-1})^{\ast} \eta)$ for each $\eta \in D((T^{-1})^{\ast})$, we have $(T^{-1})^{\ast} \eta \in D(T^{\ast})$ and $T^{\ast}(T^{-1})^{\ast} \eta = \eta$, which implies that $(T^{-1})^{\ast} \subset (T^{\ast})^{-1}$. The inverse $(T^{\ast})^{-1} \subset (T^{-1})^{\ast}$ is similarly shown. We generalize the notion of Riesz base as follows:\\
\par
{\it Definition 2.1.} {\it A sequence $\{ \phi_{n} \}$ in ${\cal H}$ is called a generalized Riesz base if there exists a densely defined closed operator $T$ in ${\cal H}$ with a densely defined inverse and there exists an ONB $\{ e_{n} \}$ in ${\cal H}$ contained in $D(T) \cap D((T^{\ast})^{-1})$ such that \\
\hspace{3mm} (i) $\phi_{n}=Te_{n}, \;\;\; n=0,1,2, \cdots,$\\
\hspace{3mm} (ii) linear span $D_{\phi}$ of $\{ \phi_{n}=Te_{n} \}$ is dense in ${\cal H}$,\\
\hspace{3mm} (iii) linear span $D_{\psi}$ of $\{ \psi_{n} \equiv (T^{-1})^{\ast} e_{n} \}$ is dense in ${\cal H}$.}\\
\par
{\it Remark:} (1) If $\{ \phi_{n} \}$ is a Riesz base, that is, both $T$ and $T^{-1}$ in Definition 2.1 are bounded, then the conditions (ii) and (iii) in Definition 2.1 hold automatically.\\
(2) If $T$ in Definition 2.1 is bounded, then (ii) holds automatically.\\
(3) If $T^{-1}$ in Definition 2.1 is bounded, then (iii) holds automatically.

In general case, though $R(T)$ and $R(T^{\ast})$ are dense in ${\cal H}$, (ii) and (iii) do not necessarily hold. As shown later, the assumption (ii) is necessary that the operators $H_{\bm{\alpha}}$, $A_{\bm{\alpha}}$ and $B_{\bm{\alpha}}$ are densely defined, and the assumption (iii) is necessary that the operators $H_{\bm{\alpha}}^{\dagger}$, $A_{\bm{\alpha}}^{\dagger}$ and $B_{\bm{\alpha}}^{\dagger}$ are densely defined.\\
Suppose that $\{ \phi_{n} \}$ is a generalized Riesz base. Then $\{ \phi_{n} \}$ is biorthogonal to $\{ \psi_{n} \}$, that is, $ \left( \phi_{n} | \psi_{m} \right) =\delta_{nm}$, $n,m=0,1,2, \cdots $. Furthermore, we have the following\\
\par
{\it Lemma 2.2.} {\it Let $T^{\ast}=U |T^{\ast}|$ be a polar decomposition of $T^{\ast}$ and $e_{n}^{\prime} \equiv U^{\ast} e_{n}$, $n=0,1,2, \cdots$. Then $\{ e_{n}^{\prime} \}$ is an ONB in ${\cal H}$, $\phi_{n}=|T^{\ast}| e_{n}^{\prime}$ and $\psi_{n}=|T^{\ast}|^{-1} e_{n}^{\prime}$, $n=0,1,2, \cdots$.} \\
\par
{\it Proof.} It is well known in \cite{reed-simon} that $U$ is a partial isometry, $|T^{\ast}|$ is a positive self-adjoint operator in ${\cal H}$ and 
\begin{eqnarray}
U^{\ast}U 
&=& \;\; {\rm projection \; on} \; \overline{R( |T^{\ast}| )} \nonumber \\
&=& \;\; {\rm projection \; on} \; \overline{R( T )}, \nonumber \\
UU^{\ast}
&=& \;\; {\rm projection \; on} \;  \overline{R( T^{\ast} )}. \nonumber
\end{eqnarray}
Since $\overline{R(T)}=\overline{R( T^{\ast} )}={\cal H}$ as shown above, it follows that $U$ is unitary.

Since $U$ is a unitary operator on ${\cal H}$, it follows that $\{ e_{n}^{\prime} \equiv U^{\ast} e_{n} \}$ is an ONB in ${\cal H}$ and
\begin{eqnarray}
\phi_{n}
&=& T e_{n} = |T^{\ast}| U^{\ast} e_{n} \nonumber \\
&=& |T^{\ast} | e_{n}^{\prime}. \nonumber
\end{eqnarray}
Furthermore, we have
\begin{eqnarray}
\psi_{n}
&=& \left( T^{-1} \right)^{\ast} e_{n} = \left( T^{\ast} \right)^{\-1} e_{n} \nonumber \\
&=& |T^{\ast}|^{-1} U^{\ast} e_{n} = |T^{\ast} |^{-1} e_{n}^{\prime}. \nonumber
\end{eqnarray}
This completes the proof. 
\hfill $\Box$
\\\\
By Lemma 2.2, we may assume that an operator $T$ in definition of generalized Riesz base is a positive self-adjoint operator without loss of generality. Let $\{ \phi_{n} \}$ be a generalized Riesz base, that is, $\phi_{n}=T e_{n}$, $n=0,1,2, \cdots$, where $\{ e_{n} \}$ is an ONB in ${\cal H}$ and $T$ is a positive self-adjoint operator in ${\cal H}$ with inverse $T^{-1}$ whose domains $D(T)$ and $D(T^{-1})$ contain $\{ e_{n} \}$. We put $\psi_{n}= T^{-1} e_{n}$, $n=0,1,2, \cdots$. Then $\{ \phi_{n} \}$ is biorthogonal to $\{ \psi_{n} \}$.

Throughout this section, let $\bm{\alpha}= \{ \alpha_{n} \}$ be any sequence of complex numbers. We define two operators
\begin{eqnarray}
H_{\bm{\alpha}}
= T \left( \sum_{n=0}^{\infty} \alpha_{n} e_{n} \otimes \bar{e}_{n} \right) T^{-1} \nonumber
\end{eqnarray}
and
\begin{eqnarray}
H_{\phi , \psi}^{\bm{\alpha}} 
&=& \sum_{n=0}^{\infty} \alpha_{n} \phi_{n} \otimes \bar{\psi}_{n}, \nonumber
\end{eqnarray}
where the tensor $ x \otimes \bar{y}$ of elements $x, \; y$ of ${\cal H}$ is defined by
\begin{eqnarray}
\left( x \otimes \bar{y} \right) \xi = \left( \xi | y \right) x, \;\;\; \xi \in {\cal H}. \nonumber
\end{eqnarray} 
It is easily shown that $ x \otimes \bar{y}$ is a bounded linear operator on ${\cal H}$ satisfying $ \left( x \otimes \bar{y} \right)^{\ast} = y \otimes \bar{x}$ and $ \| x \otimes \bar{y} \|= \| x \| \| y \|$. In details, the operators $H_{\bm{\alpha}}$ and $H_{\phi, \psi}^{\bm{\alpha}}$ are defined as follows:
\begin{equation}
\left\{
\begin{array}{cl}
 D(H_{\bm{\alpha}})
&= \left\{ \xi \in D(T^{-1}) \ ; \ T^{-1} \xi \in D \left(  \left( \sum_{n=0}^{\infty} \alpha_{n} e_{n} \otimes \bar{e}_{n} \right) \right)  \right.\\
&\hspace*{40mm} \left. {\rm and} \;  \left( \sum_{n=0}^{\infty} \alpha_{n} e_{n} \otimes \bar{e}_{n} \right) T^{-1} \xi \in D(T) \right\}  \\\\
&= \{ \xi \in D(T^{-1}) \ ; \ \sum_{n=0}^{\infty} \alpha_{n} \left( T^{-1} \xi | e_{n} \right) e_{n} \; {\rm exists \; in} \; {\cal H} \\
&\hspace*{40mm}  {\rm and \; it \; belongs \; to} \; D(T) \} \tag{2.1} \\\\
H_{\bm{\alpha}} \xi
&=  T \left( \sum_{n=0}^{\infty} \alpha_{n} e_{n} \otimes \bar{e}_{n} \right) T^{-1} \xi, \;\;\; \xi \in D(H_{\bm{\alpha}}) \\
\end{array}
\right. \nonumber
\end{equation}
\begin{equation}
\left\{
\begin{array}{cl}
D(H_{\phi , \psi}^{\bm{\alpha}} )
&= \{ \xi \in {\cal H}; \sum_{n=0}^{\infty} \alpha_{n} \left( \xi | \psi_{n} \right) \phi_{n} \; {\rm exists \; in } \; {\cal H} \}  \tag{2.2} \\\\
H_{\phi , \psi}^{\bm{\alpha}} \xi
&= \sum_{n=0}^{\infty} \alpha_{n}  \left( \xi | \psi_{n} \right) \phi_{n}, \;\;\; \xi \in D(H_{\phi , \psi}^{\bm{\alpha}}). \\
\end{array}
\right. \nonumber
\end{equation}
We first compare with operators $H_{\bm{\alpha}}$ and $H_{\phi , \psi}^{\bm{\alpha}}$. By (2.1), $\xi \in D \left( \sum_{n=0}^{\infty} \alpha_{n} \phi_{n} \otimes \bar{\psi}_{n} \right)$ if and only if
\begin{equation}
\xi \in {\cal H} \;\;\; {\rm and} \;\;\; T \left( \sum_{k=0}^{n} \alpha_{k} ( \xi | T^{-1} e_{k}) e_{k} \right) \;\; {\rm converges \; in} \; {\cal H}. \tag{2.3} 
\end{equation}
Then, in general
\begin{equation}
\xi \notin D(T^{-1}) \nonumber
\end{equation}
and
\begin{equation}
\sum_{k=0}^{n} \alpha_{k} ( \xi | T^{-1} e_{k}) e_{k}  \;\; {\rm does \; not \;converge \; in} \; {\cal H}. \nonumber
\end{equation}
Hence it follows from (2.1) that $ \xi \notin D \left( T \left( \sum_{n=0}^{\infty} \alpha_{n} e_{n} \otimes \bar{e}_{n} \right) T^{-1} \right)$ in general.\\
On the other hand, if $\xi \in D \left( T \left( \sum_{n=0}^{\infty} \alpha_{n} e_{n} \otimes \bar{e}_{n} \right) T^{-1} \right)$, then $\xi \in D(T^{-1})$ and\\ $\sum_{k=0}^{n} \alpha_{k} ( \xi | T^{-1} e_{k}) e_{k}$ converges in ${\cal H}$  and furthermore
\begin{eqnarray}
T \left( \sum_{k=0}^{n} \alpha_{k} ( \xi | T^{-1} e_{k}) e_{k} \right) \;\; {\rm converges \; weakly \; in} \; {\cal H} \nonumber
\end{eqnarray}
in the following sense:
\begin{eqnarray}
\lim_{n \rightarrow \infty}  \left( \xi \;  \left| \;\; T \left( \sum_{k=0}^{n} \alpha_{k} ( \xi | T^{-1} e_{k}) e_{k} \right. \right) \right)
= ( \xi | \eta ), \;\;\; ^{\forall} \xi \in D(T) \nonumber
\end{eqnarray}
for some $\eta \in {\cal H}$.\\
But, $T  \sum_{k=0}^{n} \alpha_{k} ( \xi | T^{-1} e_{k}) e_{k}$ does not necessary converge in ${\cal H}$. Hence, $ \xi \notin D \left( \sum_{n=0}^{\infty} \alpha_{n} \phi_{n} \otimes \bar{\psi}_{n} \right)$ in general. Thus, the operators $H_{\bm{\alpha}}$ and $H_{\phi, \psi}^{\bm{\alpha}}$ are different. But, under certain assumptions for operators $T$ and $T^{-1}$ we have the following relations between $H_{\bm{\alpha}}$ and $H_{\phi, \psi}^{\bm{\alpha}}$.\\
\par
{\it Proposition 2.3.} {\it The following statements hold.\\
(1) $H_{\phi , \psi}^{\bm{\alpha}}$ and $H_{\bm{\alpha}}$ are densely defined linear operators in ${\cal H}$.\\
(2) If $\{ \phi_{n} \}$ is a Riesz base, that is, both $T$ and $T^{-1}$ are bounded, then $H_{\phi , \psi}^{\bm{\alpha}}=H_{\bm{\alpha}}$.\\
(3) If $T$ is unbounded and $T^{-1}$ is bounded, then $H_{\phi , \psi}^{\bm{\alpha}} \subset H_{\bm{\alpha}}$.} \\
\par
{\it Proof.} (1) Since $D(H_{\phi , \psi}^{\bm{\alpha}} )$ and $ D(H_{\bm{\alpha}})$ contain $\{ \phi_n \}$ and $D_{\phi}$ is dense in ${\cal H}$, it follows that $H_{\phi , \psi}$ and $H_{\bm{\alpha}}$ are densely defined linear operators in ${\cal H}$.\\
(2) Since $T$ and $T^{-1}$ are bounded, by (2.1) and (2.2) we have $H_{\phi , \psi}^{\bm{\alpha}} = H_{\bm{\alpha}}$.\\
(3) Take an arbitrary $\xi \in D(H_{\phi, \psi}^{\bm{\alpha}} )$. Then we have 
\begin{equation}
H_{\phi, \psi}^{\bm{\alpha}} \xi
= \lim_{n \rightarrow \infty} \sum_{k=0}^{n} \alpha_{k} ( \xi | \psi_{k} ) \phi_{k} 
= \lim_{n \rightarrow \infty} T \left( \sum_{k=0}^{n} \alpha_{k} e_{k} \otimes \bar{e}_{k} \right) T^{-1} \xi. \tag{2.4}
\end{equation}
Since $T^{-1}$ is bounded, we have
\begin{eqnarray}
\lim_{n \rightarrow \infty} \left( \sum_{k=0}^{n} \alpha_{k} e_{k} \otimes \bar{e}_{k} \right) T^{-1} \xi 
&=& \left( \sum_{n=0}^{\infty} \alpha_{n} e_{n} \otimes \bar{e}_{n} \right) T^{-1} \xi \nonumber \\
&=& T^{-1} H_{\phi, \psi}^{\bm{\alpha}} \xi. \nonumber
\end{eqnarray}
Hence it follows from (2.4) and the closedness of $T$ that $\left( \sum_{n=0}^{\infty} \alpha_{n} e_{n} \otimes \bar{e}_{n} \right) T^{-1} \xi \in D(T)$ and
\begin{eqnarray}
T \left( \sum_{n=0}^{\infty} \alpha_{n} e_{n} \otimes \bar{e}_{n} \right) T^{-1} \xi
&=& T T^{-1} H_{\phi, \psi}^{\bm{\alpha}} \xi \nonumber \\
&=& H_{\phi, \psi}^{\bm{\alpha}} \xi. \nonumber
\end{eqnarray}
Thus we have $H_{\phi, \psi}^{\bm{\alpha}} \subset T \left( \sum_{n=0}^{\infty} \alpha_{n} e_{n} \otimes \bar{e}_{n} \right) T^{-1} = H_{\bm{\alpha}}$.
\hfill $\Box$
\\\\
As shown in Proposition 2.3, in case that $\{ \phi_{n} \}$ is a Riesz base, $H_{\phi , \psi}^{\bm{\alpha}}=H_{\bm{\alpha}}$. By using $H_{\bm{\alpha}}$ instead of $H_{\phi , \psi}^{\bm{\alpha}}$, all of the results in \cite{b-i-t} are almost trivial. \\
It is easily shown that $H_{\bm{\alpha}}^{\ast} \supset T^{-1} \left( \sum_{n=0}^{\infty} \bar{\alpha}_{n} e_{n} \otimes \bar{e}_{n} \right) T$. Hence we put
\begin{eqnarray}
H_{\bm{\alpha}}^{\dagger}
\equiv T^{-1} \left( \sum_{n=0}^{\infty} \bar{\alpha}_{n} e_{n} \otimes \bar{e}_{n} \right) T. \nonumber
\end{eqnarray}
Then we have the following\\
\par
{\it Proposition 2.4.} {\it The following statements hold.\\
(1) If $\{ \phi_{n} \}$ is a Riesz base, then $H_{\bm{\alpha}}^{\ast}=H_{\bm{\alpha}}^{\dagger}$.\\
(2) If $T$ is unbounded and $T^{-1}$ is bounded, then $H_{\bm{\alpha}}$ is closed, but $H_{\bm{\alpha}}^{\dagger}$ is not necessary closed.\\
(3) If $T$ is bounded and $T^{-1}$ is unbounded, then $H_{\bm{\alpha}}^{\dagger}$ is closed, but $H_{\bm{\alpha}}$ is not necessary closed.}\\
\par
{\it Proof.} (1) Take an arbitrary $\eta \in D( H_{\bm{\alpha}}^{\ast})$. Then there exists an element $\zeta$ of ${\cal H}$ such that
\begin{equation}
\left( H_{\bm{\alpha}} \xi \mid \eta \right)
=\left( T \left( \sum_{n=0}^{\infty} \alpha_{n} e_{n} \otimes \bar{e}_{n} \right) T^{-1} \xi \mid \eta \right) \nonumber\\
= (\xi \mid \zeta) \nonumber
\end{equation}
for all $\xi \in D( H_{\bm{\alpha}})$. Since $\xi \equiv T e_{n} \in D( H_{\bm{\alpha}})$ for $n=0,1,2, \cdots$, we have $ \bar{\alpha}_{n} T \eta = T \zeta$. Hence it follows since $T$ and $T^{-1}$ are bounded that
\begin{equation}
\lim_{ n \rightarrow \infty} T^{-1} \left( \sum_{k=0}^{n} \bar{\alpha}_{k} e_{k} \otimes \bar{e}_{n} \right) T \eta 
= T^{-1} \lim_{ n \rightarrow \infty}  \sum_{k=0}^{n} (T \zeta | e_{k}) e_{k} \nonumber 
= \zeta, \nonumber
\end{equation}
which implies $\eta \in D(H_{\bm{\alpha}}^{\dagger} )$ and $H_{\bm{\alpha}}^{\dagger} \eta = \zeta$. Thus we have $H_{\bm{\alpha}}^{\ast} =H_{\bm{\alpha}}^{\dagger}$.\\
(2) Take any sequence $\{ \xi_{n} \}$ in $D \left( T \left( \sum_{n=0}^{\infty} \alpha_{n} e_{n} \otimes \bar{e}_{n} \right) T^{-1} \right)$ such that $ \lim_{n \rightarrow \infty }\;  \xi_{n} = \xi$ and
\begin{equation}
\lim_{n \rightarrow \infty} T \left( \sum_{k=0}^{\infty} \alpha_{k} e_{k} \otimes \bar{e}_{k} \right) T^{-1} \xi_{n} = \eta. \tag{2.5}
\end{equation}
Since $T^{-1}$ is bounded, it follows that $\lim_{n \rightarrow \infty} \; T^{-1} \xi_{n} =T^{-1} \xi$ and 
\begin{equation}
\lim_{n \rightarrow \infty} T^{-1} T \left( \sum_{k=0}^{\infty} \alpha_{k} e_{k} \otimes \bar{e}_{k} \right) T^{-1} \xi_{n}
=\lim_{n \rightarrow \infty}  \left( \sum_{k=0}^{\infty} \alpha_{k} e_{k} \otimes \bar{e}_{k} \right) T^{-1} \xi_{n} \nonumber \\
= T^{-1} \eta, \nonumber 
\end{equation}
which implies since $\left( \sum_{n=0}^{\infty} \alpha_{n} e_{n} \otimes \bar{e}_{n} \right)$ is closed that $T^{-1} \xi \in D \left( \sum_{n=0}^{\infty} \alpha_{n} e_{n} \otimes \bar{e}_{n} \right)$ and 
\begin{equation}
\left( \sum_{n=0}^{\infty} \alpha_{n} e_{n} \otimes \bar{e}_{n} \right) T^{-1} \xi
=T^{-1} \eta. \tag{2.6}
\end{equation}
Hence we have by the closedness of $T$, $T^{-1} \eta \in D(T)$ and $TT^{-1} \eta = \eta$. Thus we have by (2.5) and (2.6)
\begin{eqnarray}
\xi \in D \left( T \left( \sum_{n=0}^{\infty} \alpha_{n} e_{n} \otimes \bar{e}_{n} \right) T^{-1} \right) \nonumber 
\end{eqnarray}
and
\begin{eqnarray}
T \left( \sum_{n=0}^{\infty} \alpha_{n} e_{n} \otimes \bar{e}_{n} \right) T^{-1} \xi
= TT^{-1} \eta = \eta. \nonumber 
\end{eqnarray}
From the above, $H_{\bm{\alpha}}$ is closed. Furthermore, by the definition of $H_{\bm{\alpha}}^{\dagger}$, $H_{\bm{\alpha}}^{\dagger}$ is not necessary closed.\\
(3) This is proved similarly to (2). 
\hfill $\Box$ \\
\section{Generalized lowering and rising operators defined by generalized Riesz bases }
In this section, we define and study generalized lowering and rising operators defined by generalized Riesz bases:
\begin{eqnarray}
A_{\bm{\alpha}}
&=& T \left( \sum_{n=0}^{\infty} \alpha_{n+1} e_{n} \otimes \bar{e}_{n+1} \right) T^{-1} , \nonumber \\
B_{\bm{\alpha}}
&=& T \left( \sum_{n=0}^{\infty} \alpha_{n+1} e_{n+1} \otimes \bar{e}_{n} \right) T^{-1} , \nonumber
\end{eqnarray}
and
\begin{eqnarray}
A_{\phi , \psi}^{\bm{\alpha}} 
&=& \sum_{n=0}^{\infty} \alpha_{n+1} \phi_{n} \otimes \bar{\psi}_{n+1}, \nonumber \\
B_{\phi , \psi}^{\bm{\alpha}} 
&=& \sum_{n=0}^{\infty} \alpha_{n+1} \phi_{n+1} \otimes \bar{\psi}_{n}. \nonumber
\end{eqnarray}
In detail, the operators $A_{\bm{\alpha}}$, $B_{\bm{\alpha}}$, $A_{\phi , \psi}^{\bm{\alpha}}$ and $B_{\phi , \psi}^{\bm{\alpha}}$ are defined as follows:
\begin{equation}
\left\{
\begin{array}{cl}
 D(A_{\bm{\alpha}})
&= \ \left\{ \xi \in D(T^{-1}) \ ; \ T^{-1} \xi \in D \left(  \left( \sum_{n=0}^{\infty} \alpha_{n+1} e_{n} \otimes \bar{e}_{n+1} \right) \right)  \right.\\
&\hspace*{40mm} \left. {\rm and} \;  \left( \sum_{n=0}^{\infty} \alpha_{n+1} e_{n} \otimes \bar{e}_{n+1} \right) T^{-1} \xi \in D(T) \right\}  \\\\
&= \ \{ \xi \in D(T^{-1}) \ ; \ \sum_{k=0}^{n} \alpha_{k+1} \left( T^{-1} \xi | e_{k+1} \right) e_{k} \; {\rm converges \; in} \; {\cal H}  \text{ as } n \rightarrow \infty \\
&\hspace*{40mm} \; {\rm and \; it \; belongs \; to} \; D(T) \}  \\\\
&= \ \{ \xi \in D(T^{-1}) \ ; \ \sum_{k=0}^{n} \alpha_{k+1} \left( T^{-1} \xi | e_{k+1} \right) e_{k} \; {\rm converges \; in} \; {\cal H} \text{ as } \; n \rightarrow \infty   \\
&\hspace*{30mm} \; {\rm and} \; T \left( \sum_{k=0}^{n} \alpha_{k+1} \left( T^{-1} \xi | e_{k+1} \right) e_{k} \right) {\rm converges \;  weakly \; in} \; {\cal H} \\
&\hspace*{30mm}  \; {\rm as} \; n \rightarrow \infty \} \\\\
A_{\bm{\alpha}} \xi
&= \  T \left( \sum_{n=0}^{\infty} \alpha_{n+1} e_{n} \otimes \bar{e}_{n+1} \right) T^{-1} \xi, \;\;\; \xi \in D(A_{\bm{\alpha}}) \\\\
\end{array}
\right. \nonumber \\ 
\end{equation}
\begin{equation}
\left\{
\begin{array}{cl}
D(B_{\bm{\alpha}})
&= \ \left\{ \xi \in D(T^{-1}) \ ; \ T^{-1} \xi \in D \left(  \left( \sum_{n=0}^{\infty} \alpha_{n+1} e_{n+1} \otimes \bar{e}_{n} \right) \right)  \right.\\
&\hspace*{40mm} \left. {\rm and} \;  \left( \sum_{n=0}^{\infty} \alpha_{n+1} e_{n+1} \otimes \bar{e}_{n} \right) T^{-1} \xi \in D(T) \right\}  \\\\
&= \ \{ \xi \in D(T^{-1}) \ ; \ \sum_{k=0}^{n} \alpha_{k+1} \left( T^{-1} \xi | e_{k} \right) e_{k+1} \; {\rm converges \; in} \; {\cal H}  \\
&\hspace*{40mm} {\rm as}\;  n \rightarrow \infty \; {\rm and \; it \; belongs \; to} \; D(T) \}  \\\\
&= \ \{ \xi \in D(T^{-1}) \ ; \ \sum_{k=0}^{n} \alpha_{k+1} \left( T^{-1} \xi | e_{k} \right) e_{k+1} \; {\rm converges \; in} \; {\cal H} \text{ as } \; n \rightarrow \infty   \\
& \hspace*{32mm} \; {\rm and} \; T \left( \sum_{k=0}^{n} \alpha_{k+1} \left( T^{-1} \xi | e_{k} \right) e_{k+1} \right) {\rm converges \;  weakly \; in} \; {\cal H} \\
& \hspace*{32mm}  \; {\rm as} \; n \rightarrow \infty \} \\\\
B_{\bm{\alpha}} \xi
&= \ T \left( \sum_{n=0}^{\infty} \alpha_{n+1} e_{n+1} \otimes \bar{e}_{n} \right) T^{-1} \xi, \;\;\; \xi \in D(B_{\bm{\alpha}}) \\
\end{array}
\right. \nonumber
\end{equation}
and
\begin{equation}
\left\{
\begin{array}{cl}
 D(A_{\phi , \psi}^{\bm{\alpha}} )
&= \ \{ \xi \in {\cal H}; \sum_{n=0}^{\infty} \alpha_{n+1} \left( \xi | \psi_{n+1} \right) \phi_{n} \; {\rm exists \; in } \; {\cal H} \}  \\\\
A_{\phi , \psi}^{\bm{\alpha}} \xi
&= \ \sum_{n=0}^{\infty} \alpha_{n+1}  \left( \xi | \psi_{n+1} \right) \phi_{n}, \;\;\; \xi \in D(A_{\phi , \psi}^{\bm{\alpha}}) \\
\end{array}
\right. \nonumber
\end{equation}
\begin{equation}
\left\{
\begin{array}{cl}
D(B_{\phi , \psi}^{\bm{\alpha}} )
&= \ \{ \xi \in {\cal H}; \sum_{n=0}^{\infty} \alpha_{n+1} \left( \xi | \psi_{n} \right) \phi_{n+1} \; {\rm exists \; in } \; {\cal H} \}   \\\\
B_{\phi , \psi}^{\bm{\alpha}} \xi
&= \ \sum_{n=0}^{\infty} \alpha_{n+1}  \left( \xi | \psi_{n} \right) \phi_{n+1}, \;\;\; \xi \in D(B_{\phi , \psi}^{\bm{\alpha}}). \\
\end{array}
\right. \nonumber
\end{equation}\\
As shown in case of $H_{\bm{\alpha}}$ and $H_{\phi,\psi}^{\bm{\alpha}}$ in Section 2, $A_{\bm{\alpha}}$ and $A_{\phi,\psi}^{\bm{\alpha}}$ are different and $B_{\bm{\alpha}}$ and $B_{\phi,\psi}^{\bm{\alpha}}$ are also different. But, under some conditions for $T$ and $T^{-1}$ we have following\\
\par
{\it Proposition 3.1.} {\it The following statements hold.\\
(1)  $A_{\bm{\alpha}}$, $B_{\bm{\alpha}}$, $A_{\phi , \psi}^{\bm{\alpha}}$ and $B_{\phi , \psi}^{\bm{\alpha}}$ are densely defined linear operators in ${\cal H}$.\\
(2) If $\{ \phi_{n} \}$ is a Riesz base, then $A_{\bm{\alpha}}= A_{\phi , \psi}^{\bm{\alpha}}$ and $B_{\bm{\alpha}}=B_{\phi , \psi}^{\bm{\alpha}}$.\\
(3) If $T$ is unbounded and $T^{-1}$ is bounded, then $A_{\phi , \psi}^{\bm{\alpha}} \subset A_{\bm{\alpha}}$ and $B_{\phi , \psi}^{\bm{\alpha}} \subset B_{\bm{\alpha}}$.} \\
\par
{\it Proof.} These are proved similarly to Proposition 2.3. 
\hfill $\Box$ \\\\
To investigate the operators $A_{\bm{\alpha}}$ and $B_{\bm{\alpha}}$ we define operators $A_{\bm{\alpha}}^{\dagger}$ and $B_{\bm{\alpha}}^{\dagger}$ by
\begin{eqnarray}
A_{\bm{\alpha}}^{\dagger}
&= T^{-1} \left( \displaystyle \sum_{n=0}^{\infty} \bar{\alpha}_{n+1} e_{n+1} \otimes \bar{e}_{n} \right) T, \nonumber \\
B_{\bm{\alpha}}^{\dagger}
&= T^{-1} \left( \displaystyle\sum_{n=0}^{\infty} \bar{\alpha}_{n+1} e_{n} \otimes \bar{e}_{n+1} \right) T. \nonumber
\end{eqnarray}
Clearly, $A_{\bm{\alpha}}^{\ast} \supset A_{\bm{\alpha}}^{\dagger}$ and $B_{\bm{\alpha}}^{\ast} \supset B_{\bm{\alpha}}^{\dagger}$ , furthermore we have following\\
\par
{\it Proposition 3.2.} {\it The following statements hold.\\
(1) If $\{ \phi_{n} \}$ is a Riesz base, then $A_{\bm{\alpha}}^{\ast}=A_{\bm{\alpha}}^{\dagger}$ and $B_{\bm{\alpha}}^{\ast} = B_{\bm{\alpha}}^{\dagger}$.\\
(2) If $T$ is unbounded and $T^{-1}$ is bounded, then $A_{\bm{\alpha}}$ and $B_{\bm{\alpha}}$ are closed, but $A_{\bm{\alpha}}^{\dagger}$ and $B_{\bm{\alpha}}^{\dagger}$ are not necessary closed.\\
(3)If $T$ is bounded and $T^{-1}$ is unbounded, then $A_{\bm{\alpha}}^{\dagger}$ and $B_{\bm{\alpha}}^{\dagger}$ are closed, but $A_{\bm{\alpha}}$ and $B_{\bm{\alpha}}$ are not necessary closed.}\\
\par
{\it Proof.} The statements (1), (2) and (3) are proved similarly to (1), (2) and (3) in Proposition 2.4, respectively. \\\\
Since the relevance of these operators $A_{\bm{\alpha}}$ and $B_{\bm{\alpha}}$ relies essentially on their products, we give the following\\
\par
{\it Theorem 3.3.} {\it The following statements hold.\\
(1) $A_{\bm{\alpha}}B_{\bm{\alpha}} \subset T \left( \sum_{n=0}^{\infty} \alpha_{n+1} ^{2}  e_{n} \otimes \bar{e}_{n} \right) T^{-1}$.\\
(2) $B_{\bm{\alpha}}A_{\bm{\alpha}} \subset T \left( \sum_{n=0}^{\infty}  \alpha_{n+1}^{2} e_{n+1} \otimes \bar{e}_{n+1} \right) T^{-1}$.\\
(3) $A_{\bm{\alpha}}B_{\bm{\alpha}}-B_{\bm{\alpha}}A_{\bm{\alpha}} \subset T \left( \sum_{n=0}^{\infty}  \alpha_{n+1} ^{2} e_{n} \otimes \bar{e}_{n} - \sum_{n=0}^{\infty} \alpha_{n+1} ^{2} e_{n+1} \otimes \bar{e}_{n+1} \right) T^{-1}$.} \\
\par
{\it Proof.}\\
(1) By the definition of the operators $A_{\bm{\alpha}}$, $B_{\bm{\alpha}}$, we have
\begin{eqnarray}
 D(A_{\bm{\alpha}} B_{\bm{\alpha}})
&=& \left\{ \xi \in D(T^{-1}); T^{-1} \xi \in D \left(  \left( \sum_{n=0}^{\infty} \alpha_{n+1} e_{n+1} \otimes \bar{e}_{n} \right) \right)  \right. \nonumber \\ \nonumber \\
&&\hspace*{30mm} {\rm and} \;  \left( \sum_{n=0}^{\infty} \alpha_{n+1} e_{n+1} \otimes \bar{e}_{n} \right) T^{-1} \xi \in D(T) \nonumber \\
&&\hspace*{30mm}\left. {\rm and} \;  \left( \sum_{n=0}^{\infty}  \alpha_{n+1} ^{2} e_{n} \otimes \bar{e}_{n} \right) T^{-1} \xi \in D(T) \right\}  \nonumber \\
&\subset& \left\{ \xi \in D(T^{-1});   \left( \sum_{n=0}^{\infty} \alpha_{n+1}^{2} e_{n} \otimes \bar{e}_{n} \right) T^{-1} \xi \in D(T) \right\}  \nonumber \\
&=& D \left( T  \left( \sum_{n=0}^{\infty} \alpha_{n+1}^{2} e_{n} \otimes \bar{e}_{n} \right) T^{-1} \right) \nonumber
\end{eqnarray}
and
\begin{eqnarray}
A_{\bm{\alpha}}B_{\bm{\alpha}} \xi
=  T \left( \sum_{n=0}^{\infty}  \alpha_{n+1}^{2} e_{n} \otimes \bar{e}_{n} \right) T^{-1} \xi , \;\;\; ^{\forall} \xi \in D( A_{\bm{\alpha}}B_{\bm{\alpha}}). \nonumber
\end{eqnarray}
Hence,
\begin{eqnarray}
A_{\bm{\alpha}}B_{\bm{\alpha}} \subset T \left( \sum_{n=0}^{\infty} \alpha_{n+1}^{2} e_{n} \otimes \bar{e}_{n} \right) T^{-1}. \nonumber
\end{eqnarray}
(2) By the definition of the operators $A_{\bm{\alpha}}$, $B_{\bm{\alpha}}$, we have
\begin{eqnarray}
 D(B_{\bm{\alpha}} A_{\bm{\alpha}})
&=& \left\{ \xi \in D(T^{-1}); T^{-1} \xi \in D \left(  \left( \sum_{n=0}^{\infty} \alpha_{n+1} e_{n} \otimes \bar{e}_{n+1} \right) \right)  \right. \nonumber \\ \nonumber \\
&&\hspace*{30mm}  {\rm and} \;  \left( \sum_{n=0}^{\infty} \alpha_{n+1} e_{n} \otimes \bar{e}_{n+1} \right) T^{-1} \xi \in D(T) \nonumber \\
&&\hspace*{30mm} \left. {\rm and} \;  \left( \sum_{n=0}^{\infty} \alpha_{n+1}^{2} e_{n+1} \otimes \bar{e}_{n+1} \right) T^{-1} \xi \in D(T) \right\}  \nonumber \\
&\subset& \left\{ \xi \in D(T^{-1}); \left( \sum_{n=0}^{\infty} \alpha_{n+1}^{2} e_{n+1} \otimes \bar{e}_{n+1} \right) T^{-1} \xi \in D \left(  T \right)  \right\} \nonumber \\ \nonumber \\
&=& D \left( T  \left( \sum_{n=0}^{\infty} \alpha_{n+1}^{2} e_{n+1} \otimes \bar{e}_{n+1} \right) T^{-1} \right) \nonumber
\end{eqnarray}
and
\begin{eqnarray}
B_{\bm{\alpha}}A_{\bm{\alpha}} \xi
=  T \left( \sum_{n=0}^{\infty} \alpha_{n+1}^{2} e_{n+1} \otimes \bar{e}_{n+1} \right) T^{-1} \xi , \;\;\; ^{\forall} \xi \in D( B_{\bm{\alpha}}A_{\bm{\alpha}}). \nonumber
\end{eqnarray}
Hence,
\begin{eqnarray}
B_{\bm{\alpha}}A_{\bm{\alpha}} \subset T \left( \sum_{n=0}^{\infty} \alpha_{n+1}^{2} e_{n+1} \otimes \bar{e}_{n+1} \right) T^{-1}. \nonumber
\end{eqnarray}
(3) This follows from (1) and (2).\hfill $\Box$ \\\\
Putting $ \alpha_{n}= \sqrt{n}, \; n=0,1,2, \cdots $, we denote $A_{\bm{\alpha}}$ and $B_{\bm{\alpha}}$ by $A$ and $B$, respectively. By Theorem 3.3 (3), we have the following\\
\par
{\it Corollary 3.3.} $AB-BA \subset I.$\\
\par
{\it Example 3.4.} Let $\pi_{s}$ be the Schr$\ddot{{\rm o}}$dinger representation of the CCR-algebra generated by self-adjoint elements $p$ and $q$ satisfying the Heisenberg commutation relation $pq-qp= -i 1$ defined by
\begin{eqnarray}
 \left( \pi_{s} (p) f \right) (t)
&=& -i \frac{d}{dt} f(t), \nonumber \\
\left( \pi_{s} (p) g \right) (t)
&=& t f(t), \;\;\; f \in S( \bm{R}), \nonumber
\end{eqnarray}
where $S(\bm{R})$ is the space of infinitely differentiable rapidly decreasing functions on $\bm{R}$ 
, see \cite{powers71},\cite{powers82}. Let $\{ f_{n} \}$ be an ONB in the Hilbert space $L^{2}(\bm{R})$ contained in $S(\bm{R})$ defined by
\begin{eqnarray}
f_{n}(t)
= \pi^{- \frac{1}{4}} \left( 2^{n} n ! \right)^{- \frac{1}{2}} \left( t- \frac{d}{dt} \right)^{n} e^{- \frac{t^{2}}{2}} ,  \;\;\; n=0,1,2, \cdots . \nonumber
\end{eqnarray}
Then, putting $a= \frac{1}{\sqrt{2}} (q+ip)$, the following
\begin{eqnarray}
\pi_{s} (aa^{\ast}) f_{n}
&=& (n+1) f_{n} , \;\;\; n=0,1,2 \cdots , \nonumber \\
\pi_{s}(a) f_{n}
&=& \left\{
\begin{array}{cccc}
 0 &  , & &n=0,   \\
 \sqrt{n} f_{n-1} &  , && n=1 ,2, \cdots  , \\
\end{array}
\right. \nonumber \\
\pi_{s} (a^{\ast}) f_{n}
&=& \sqrt{n+1} f_{n+1}, \;\;\; n=0,1,2, \cdots , \nonumber 
\end{eqnarray}
hold, and so the number operator $H_{0}$, the lowering operator $A_{0}$ and the rising operator $A_{0}^{\ast}$ as follows:
\begin{eqnarray}
H_{0}
&=& \sum_{n=0}^{\infty} (n+1) \; f_{n} \otimes \bar{f}_{n}, \nonumber \\
A_{0}
&=& \sum_{n=0}^{\infty} \sqrt{n+1} \; f_{n} \otimes \bar{f}_{n+1}, \nonumber \\
A_{0}^{\ast}
&=& \sum_{n=0}^{\infty} \sqrt{n+1} \; f_{n+1} \otimes \bar{f}_{n}. \nonumber 
\end{eqnarray}
For any positive self-adjoint operator $T$ in $L^{2}(\bm{R})$ with inverse whose domains $D(T)$ and $D(T^{-1})$ contain $\{ f_{n} \}$, non self-adjoint hamiltonian $H \equiv TH_{0}T^{-1}$, the generalized lowering operator $A \equiv TA_{0}T^{-1}$ and  the generalized rising operator $B \equiv TA_{0}^{\ast}T^{-1}$ are considered as stated here. In particular, if the domain of $T$ contains $S(\bm{R})$, then the domains of $H$, $A$ and $B$ contain $TS(\bm{R})$, and if the domain of $T^{-1}$ contain $S(\bm{R})$, then the domains of $H^{\dagger}$, $A^{\dagger}$ and $B^{\dagger}$ contain $T^{-1}S( \bm{R})$. Indeed, this follows since $S(\bm{R})$ is an invariant subspace for $H_{0}$, $A_{0}$ and $B_{0}$. A subspace ${\cal D}$ is called invariant for an operator $X$ if ${\cal D} \subset D(X)$ and $X {\cal D} \subset {\cal D}$. Furthermore, if $S(\bm{R})$ is invariant for $T$ and $T^{-1}$, equivalently $TS(\bm{R})=S(\bm{R})$, then $S(\bm{R})$ is an invariant subspace for $H$, $A$ and $B$ and their adjoints $H^{\dagger}$, $A^{\dagger}$ and $B^{\dagger}$. Hence, a $\ast$-algebra generated by the restrictions of these operators to $S(\bm{R})$ is considered. In this case, we may easily treat with the operators $H, \; A$ and $B$.\\
We think that these operators may be useful for studies of  ${\it quasi}$-${\it hermitian \; quantum \; mechanics}$ and its relatives.

\ \\
Graduate School of Mathematics, Kyushu University, 744 Motooka, Nishi-ku, Fukuoka 819-0395, Japan
\\
h-inoue@math.kyushu-u.ac.jp, 

\ \\
Department of Applied Mathematics, Fukuoka University, Fukuoka 814-0180, Japan\\
mayumi@fukuoka-u.ac.jp, \\

\begin{thebibliography}{99}
\bibitem{reed-simon}
M. Reed and B. Simon", Methods of modern Mathmatical Physics, vol. I,
Academic Press, New York, 1980


\bibitem{b-i-t}
F. Bagarello and A. Inoue and C. Trapani,
Non-self-adjoint hamiltonians defined by Riesz bases,
J. Math. Phys., {\bf 55}(2014), 033501


\bibitem{bagarello13}
F. Bagarello,
More mathematics for pseudo-bosons,
J. Math. Phys., {\bf 58}(2013), 063512


\bibitem{bagarello10}
F. Bagarello,
Pseudobosons, Riesz bases, and coherent states,
J. Math. Phys., {\bf 51}(2010), 023531

\bibitem{powers71}
R.T. Powers ,
Self-adjoint algebras of unbounded operators, 
Comm. Math. Phys., {\bf 21}(1971), 85-124


\bibitem{powers82}
 R.T. Powers, Algebras of Unbounded operators, Proc. Sympos. Pure Math., {\bf 38}(1982), 389-406

\bibitem{mostafazadeh}
 A. Mostafazadeh, 
 Pseudo-Hermitian representatoion of Quantum Mechanics , Int. J. Geom. Methods Mod. Phys., {\bf 7}(2010), 1191-1306 ,


\bibitem{bagarello11}
 F. Bagarello, (Regular) pseudo-bosons versus bosons, Int. J. Geom. Methods Mod. Phys. A, {\bf 44}(2011), 015205

\bibitem{bagarello2013}
 F. Bagarello ,
 From self to non self-adjoint harmonic oscillators: physical consequences and mathematical pitfalls, Phys. Rev. A, {\bf 88}(2013), 032120

\end{thebibliography}
\end{document}